\documentclass[twocolumn,prl,showpacs,amsmath,amstex,amssymb,mathfonts,superscriptaddress]{revtex4-1}
\pdfoutput=1
\usepackage{color}
\usepackage{verbatim}
\usepackage{amsmath}
\usepackage{amssymb}
\usepackage{amsthm}
\usepackage{amsfonts}
\usepackage{listings}
\usepackage{enumerate}
\usepackage{latexsym}
\usepackage{bm}
\usepackage{graphicx}
\usepackage{hyperref}
\hypersetup{
    unicode=false,          
    pdftoolbar=false,        
    pdfmenubar=false,        
    pdffitwindow=false,     
    pdfstartview={FitH},    
    pdftitle={Universal slow growth of entanglement in interacting strongly disordered systems},    
    pdfauthor={M. Serbyn, Z. Papic, and D. Abanin},     
    pdfsubject={},   
    pdfcreator={},   
    pdfproducer={}, 
    pdfkeywords={} {} {}, 
    pdfnewwindow=true,      
    colorlinks=true,       
    linkcolor=[rgb]{0.09, 0.09, 0.5},          
    citecolor=[rgb]{0.09, 0.09, 0.5},        
    filecolor=magenta,      
    urlcolor=[rgb]{0.09, 0.09, 0.5}           
}
\newcommand{\be}{\begin{equation}}
\newcommand{\ee}{\end{equation}}
\newcommand{\bea}{\begin{eqnarray}}
\newcommand{\eea}{\end{eqnarray}}

\newcommand{\la}{\langle}
\newcommand{\ra}{\rangle}

\renewcommand{\phi}{\varphi}
\renewcommand{\epsilon}{\varepsilon}

\newcommand{\RS}{{\cal R}}
\newcommand{\LS}{{\cal L}}
\newcommand{\xd}{x}
\newcommand{\Ss}{L}
\newcommand{\Ww}{W}

\begin{document}
\title{Universal Slow Growth of Entanglement in Interacting Strongly Disordered Systems}
\author{Maksym Serbyn} 
\affiliation{Department of Physics, Massachusetts Institute of Technology, Cambridge, MA 02138, USA}
\author{Z. Papi\'c}
\affiliation{Department of Electrical Engineering, Princeton University, Princeton, NJ 08544, USA}
\author{Dmitry A. Abanin}
\affiliation{Perimeter Institute for Theoretical Physics, Waterloo, ON N2L 2Y5, Canada}
\affiliation{Institute for Quantum Computing, Waterloo, ON N2L 3G1, Canada}

\date{\today}
\begin{abstract}
Recent numerical work by Bardarson~\emph{et.~al.}~\cite{Moore12} revealed a slow, logarithmic in time, growth of the entanglement entropy for initial product states in a putative many-body localized phase. We show that this surprising phenomenon results from the dephasing due to exponentially small interaction-induced corrections to the eigenenergies of different states. 
For weak interactions, we find that the entanglement entropy grows as $\xi \ln (Vt/\hbar)$, where $V$ is the interaction strength, and $\xi$ is the single-particle localization length. The saturated value of the entanglement entropy at long times is determined by the participation ratios of the initial state over the eigenstates of the subsystem. Our work shows that the logarithmic entanglement growth is a universal phenomenon characteristic of the many-body localized phase in any number of spatial dimensions, and reveals a broad hierarchy of dephasing time scales present in such a phase.
\end{abstract}
\maketitle

{\it Introduction.---}While it is well known that arbitrarily weak disorder localizes all single-particle quantum-mechanical states in one and two dimensions, the effect of a disorder potential on the states of interacting systems largely remains an open problem. References~\cite{Basko06,Mirlin05} conjectured that localization in a many-body system survives in the presence of weak interactions. When the strength of the interactions is increased, at some critical value a transition to the delocalized phase---a ``many-body localization" transition---takes place, as observed in the numerical simulations~\cite{OganesyanHuse,Znidaric08,Monthus10, Berkelbach10, PalHuse,GogolinMueller,SilvaPRB,RigolPRL,Luca11,Cuevas12,Luca13}.  

An important challenge is to understand the physical properties of the many-body localized (MBL) phase. Recent work~\cite{Moore12} (see also Ref.~\cite{Znidaric08}) revealed that even very weak interactions dramatically change the growth of entanglement of nonequilibrium many-body states. The authors of Ref.~\cite{Moore12} studied the time evolution of product states in a 1D disordered $XXZ$ spin chain. In the absence of interactions, such states maintain a low degree of entanglement upon evolution, and the entanglement entropy $S_{\rm ent}$ obeys an area law. In contrast, in the presence of interactions the states showed a slow, logarithmic in time, growth of $S_{\rm ent}$ (here and below we use ``entanglement" and ``entanglement entropy" interchangeably). The saturated value of $S_{\rm ent}$ was found to vary approximately linearly with system size, and remained well below the maximum possible value~\cite{Moore12,Chiara06,Igloi12}.

In this Letter, we identify a mechanism that underlies the logarithmic growth of entanglement in interacting MBL states. The key observation is that although very weak interactions have a small effect on the MBL eigenstates, they nevertheless induce small corrections to their energies, which ultimately lead to the dephasing between different eigenstates at long time scales. We argue that this gives rise to a logarithmic growth of $S_{\rm ent}$ with time for a broad class of initial states that are a product of states in the two subsystems, a special example of which was considered in Ref.~\cite{Moore12}. 

For weak interactions, our mechanism leads to the following predictions regarding entanglement growth as a function of the system's parameters: (i) entropy grows as $S_{\rm ent}(t)\propto \xi {\log (Vt/\hbar)}$, where $V$ is the interaction strength and $\xi$ is the single-particle localization length; (ii) the saturation value of $S_{\rm ent}$ is of the order of the ``diagonal entropy" $S_{\rm diag}$~\cite{Polkovnikov} of the given initial state. Diagonal entropy is determined by the participation ratios of initial states in the basis of eigenstates of the system for $V=0$. We also illustrate these predictions with numerical simulations of finite systems, in particular by constructing examples of initial states for which the saturated $S_{\rm ent}$ is equal to $S_{\rm diag}$.

\begin{figure}[t]
\begin{center}
\includegraphics[width=\columnwidth]{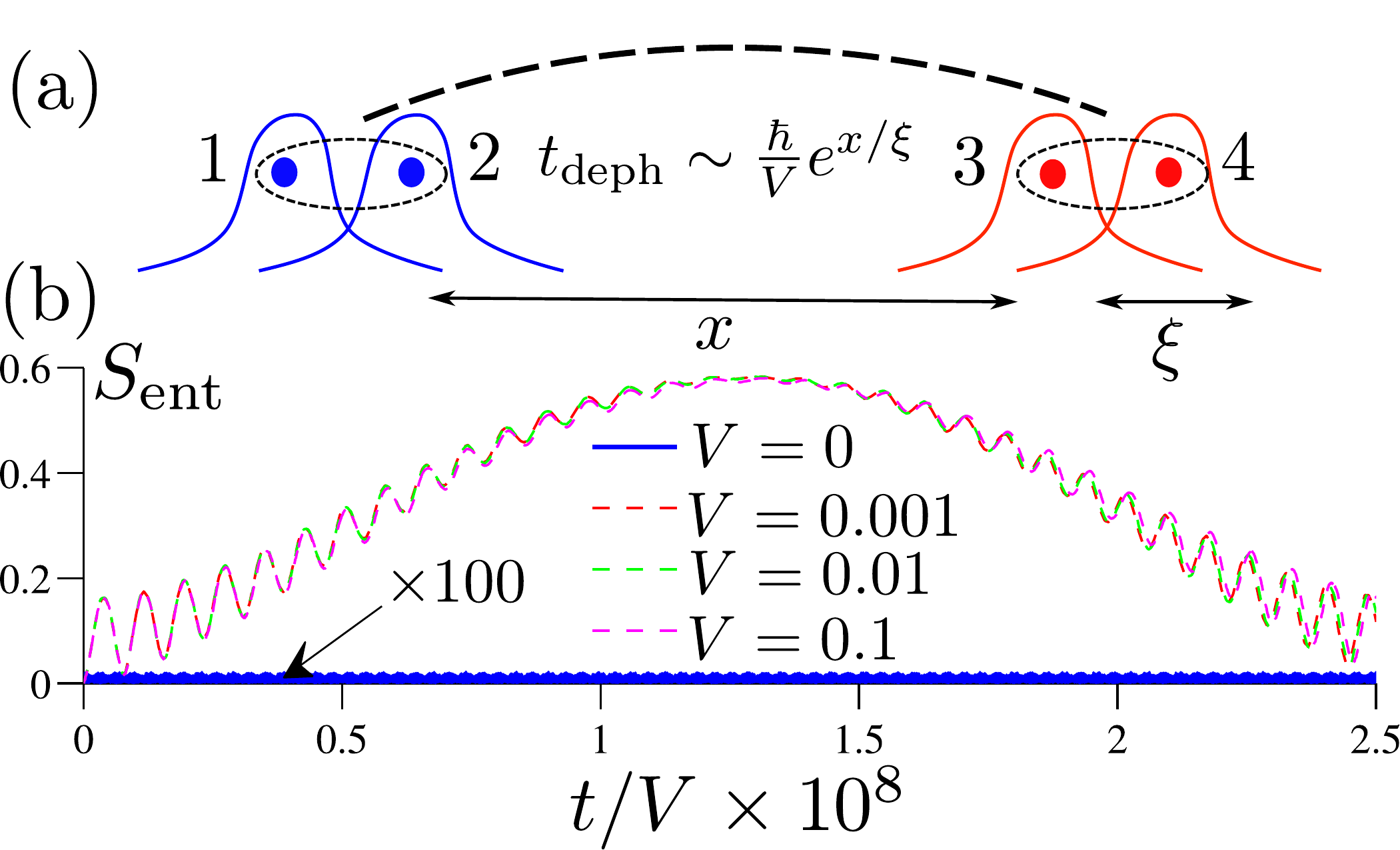}
\caption{ 
\label{Fig:2part} (Color online)
(a)~Generation of entanglement  between two remote particles, each prepared in an equal superposition of two eigenstates. Exponentially small overlap of the orbitals leads to the dephasing time growing exponentially with distance.
(b)~$S_{\rm ent}$ as a function of time for a given realization of disorder and different interaction strengths. When  $V=0$, values of $S_\text{ent}\sim 10^{-4}$ and remains small at all times. For $V \neq 0$, $S_\text{ent}(t)$  collapse on a single curve when time is scaled by $1/V$. System size is $L=10$ sites, and disorder strength is $\Ww = 6$. 
}
\end{center}
\end{figure}

{\it Model.---}Without loss of generality, we consider a 1D lattice model of fermions with on-site disorder and  nearest-neighbor interactions
\be\label{eq:hamiltonian}
H=J\sum_{\la ij\ra} c_i^\dagger c_j +\sum_i W_i \hat n_i +V\sum_{\la ij \ra} \hat n_i \hat n_j, \,\, 
\hat n_i=c_i^\dagger c_i, \,\,  
\ee
where $i,j=1,\ldots,N$, and $\la ij \ra$ denotes nearest neighbors. This model is equivalent to the random-field $XXZ$ spin chain~\cite{Moore12}. From our discussion below, it will become apparent that the logarithmic growth of entanglement in MBL systems is a robust phenomenon which does not depend on the dimensionality or the microscopic details of the system. 

We will focus mostly on the regime of weak interactions for which the logarithmic growth of $S_{\rm ent}$ found in Ref.~\cite{Moore12}  is perhaps the most striking. In the absence of interactions, $V=0$, disorder localizes the single-particle states, with localization length~$\xi$, and the many-body eigenstates are simply states in which a certain number of single-particle orbitals is occupied. Interactions that are much weaker compared to the typical level spacing $\sim 1/\xi$ do not significantly modify the many-body eigenstates. We have explicitly verified this statement for small systems, and assume it holds in general. However, even though the eigenstates are not strongly affected by the interactions, their energies are modified. If we fix the positions of all particles, except for a pair of particles situated at a distance $\xd \gg \xi$ away from each other, the interaction energy of this pair is $\sim Ve^{-\xd/\xi}$, and the corresponding dephasing time is $t_{\rm deph}\sim \hbar e^{x/\xi}/V$.  This gives rise to a hierarchy of dephasing time scales present in the problem, ranging from the fastest $t_{\rm min}=\hbar/V$ to the slowest $t_{\rm max}=t_{\rm min}e^{\Ss/\xi}$, where $\Ss$ is the system size. 

Generally, the product initial states considered in Ref.~\cite{Moore12}, as well as the initial states of other kinds considered below, are a superposition of many eigenstates. The interactions introduce a slow dephasing between different states, and effectively generate entanglement between different remote parts of the system. A subsystem of size $\xd$ becomes nearly maximally entangled with the rest of the system after an exponentially long time $t_{\rm deph}(\xd)\sim \hbar e^{\xd/\xi}/V$;  thus, the bipartite $S_{\rm ent}$ will increase logarithmically in time. 

{\it Two particles.---}Let us start with a simple example which demonstrates that the slow growth of entanglement occurs for just two particles. 
Consider two distant particles prepared in an equal-weight superposition of two neighboring localized orbitals, $|\Psi_0\ra=\frac{1}{2}(c_1^\dagger+c_2^\dagger)(c_3^\dagger+c_4^\dagger)|0\ra$, where $c_i^\dagger$ creates an eigenstate localized near site $i$. We assume that the distance between the support of the wave functions 1,2 and 3,4 is large, $\xd \gg \xi$ (see Fig.~\ref{Fig:2part}). 

In the absence of interactions, no entanglement is generated during time evolution. Interactions, however, introduce a correction to the energy of the state $ |\alpha\beta\ra=c_{\alpha}^\dagger c_\beta^\dagger|0\ra,$ where $\alpha=1,2$, $\beta=3,4$. In the leading order of perturbation theory, the energy of this state is given by $E_{\alpha\beta}=\epsilon_\alpha+\epsilon_\beta+\delta E_{\alpha\beta}$, where $\epsilon_\alpha,\epsilon_\beta$ are the single-particle energies, and the last term $\delta E_{\alpha\beta}= C_{\alpha\beta} V e^{-\xd/\xi}$ is due to the interactions, $C_{\alpha\beta}$ being a constant which depends only algebraically on $x$. 

The time-evolved state is given by $|\Psi(t)\ra=\frac{1}{2}\sum_{\alpha,\beta}   \exp(-iE_{\alpha\beta}t) |\alpha\beta\rangle$, 
and the reduced density matrix for the first particle reads
\be\label{eq:reduced}
\hat\rho_L=   \frac{1}{2}
\left( \begin{array}{cc} 1  & F(t)/2 \\ F^*(t)/2   & 1 \end{array} \right)  ,  
\ee
where $F(t)=e^{-i\Omega t}(1+e^{-i\!\: \delta \Omega\!\: t}), \, \delta\Omega=\delta E_{14}-\delta E_{24}-\delta E_{13}+\delta E_{23}$, and $\Omega = \epsilon_1-\epsilon_2+\delta E_{13}-\delta E_{23}$. The eigenstates of $\hat\rho_L$ therefore oscillate with a very long period $T=2\pi/\delta \Omega \sim ({\hbar}/{V})e^{\xd/\xi}$. At times $t=(2n+1)\pi/\delta \Omega$, the off-diagonal elements vanish, and the eigenvalues become equal to $1/2$. At these times, the particles become maximally entangled with $S_{\rm ent}=\ln 2$.  Figure~\ref{Fig:2part} demonstrates that even weak interactions lead to the entanglement of the order of $S_\text{ent} \approx \ln 2$, and the rate of entanglement change is inversely proportional to the interaction strength. In Fig.~\ref{Fig:2part}, particles are in a superposition of states which are not the exact eigenstates; hence, the maximum value of $S_\text{ent}$ is slightly below $\ln 2 \approx 0.69$. Note that no disorder or time averaging is used.

{\it General case.---}Turning to the general many-body case, let us divide the system into two parts, $\LS$ and $\RS$, labeling the single-particle orbitals that are localized dominantly in $\LS$ by index $\alpha_n$, and those residing in $\RS$ by $\beta_n$. There may be some ambiguity for the state residing near the boundary between $\LS$ and $\RS$, but we will be interested in systems of size $\Ss \gg \xi$, for which the boundary effects are not very important. 

We consider initial states that are products of some superposition of states with definite numbers of particles in $\LS$ and $\RS$: 
\be\label{eq:initial_state}
|\Psi(t=0)\ra= \sum_{\{\alpha\}\in \LS} A_{\{ \alpha \}} |\alpha_1 ...\alpha_K \ra \times \sum_{\{\beta\}\in \RS} B_{\{ \beta \}} |\beta_1 ...\beta_M\ra.
\ee
Coefficients $A,B$ are chosen such that $\Psi$ is normalized. 

Neglecting the change to the eigenstate due to interactions, the reduced density matrix for $\LS$ after time evolution reads  
$\hat\rho_{\LS} =\sum_{\alpha ,\alpha' } \rho_{\alpha  \alpha' } | \alpha\ra \la \alpha'|$, where
$\rho_{\alpha \alpha' } = A_{\alpha}A^*_{ \alpha' } \sum_{\beta } |B_{\beta }|^2  e^{i(E_{\alpha'\beta} -E_{\alpha\beta})t}$, and we have used a short-hand notation $\alpha\equiv \{\alpha \}, \beta\equiv \{\beta \}$. 
It is convenient to define $A_{\alpha}(t) = A_\alpha e^{-iE_\alpha t}$, where $E_\alpha$ is the energy of the $|\alpha\ra$ state for the isolated $\LS$ subsystem. Assuming that $|\alpha\ra\times |\beta\ra$ remains an eigenstate (this may not be true near the boundary, but the boundary effect is not important for entanglement growth, at least in large systems), the above equation, written in terms of coefficients $A_\alpha(t)$, preserves the same form, except the energies $E_{\alpha \beta}$ should be substituted by the interaction energy $\delta E_{\alpha\beta}$ between particles in the $\LS$ and $\RS$ subsystems. For particles that reside far away from the boundary, this correction can be calculated in perturbation theory. 

The energy difference $\delta E_{\alpha' \beta}-\delta E_{\alpha \beta }$ that enters the off-diagonal elements of $\hat \rho_{\LS}$, to the leading order, is proportional to $Ve^{-\xd/\xi}$. Here $\xd$ is the minimum distance between a particle in $\LS$, the position of which is different in states $\alpha$ and $\alpha'$, and the particles in $\RS$. However, it also contains many smaller contributions, which arise due to the interaction between more distant pairs of particles. Thus, the off-diagonal elements oscillate at a number of very different, incommensurate frequencies. 

The interaction energy leads to dephasing, which decreases the off-diagonal elements of $\hat \rho_{\LS}$, thus generating entanglement. Effectively, at times $t(\xd)\sim t_\text{min} e^{\xd/\xi}$ the degrees of freedom within a distance $x(t)\sim  \ln (t/t_\text{min})$ from the boundary between $\LS$ and $\RS$ are affected by dephasing, while states that differ only in the positions of particles further away from the boundary are still phase coherent. This generates the entropy
\begin{equation}\label{eq:sinf}
S_{\rm ent} (t)=  CS_{\rm diag}, \,\, S_{\rm diag}=-\sum P_i (\xd) \ln P_i(\xd), 
\end{equation}
where $P_i(\xd)$ are the probabilities of different states $|\alpha\ra$ in a segment of size $\xd$, calculated from the wave function of the initial state.  Quantity $S_{\rm diag}$ is the diagonal entropy --- a maximum achievable entropy for a given initial state, assuming that interactions do not change the eigenstates. $S_{\rm ent}$ is expected to be smaller than $S_{\rm diag}$ by a factor $C\lesssim 1$; the precise value of this prefactor is nonuniversal, and depends on the preparation of the initial state. 
In the long-time limit, assuming that $\RS \gg \LS$ and the initial state is a superposition of many different states, we expect the off-diagonal elements to become very small such that the entanglement entropy approaches its maximum value with $C\to 1$. 

Since for initial product states $S_{\rm diag}$ is proportional to the subsystem size, entanglement grows logarithmically:
\be\label{eq:Sent}
S_{\rm ent}(t)\propto \xi \log(Vt/\hbar). 
\ee
We emphasize that our argument does not rely on averaging, and therefore entanglement grows according to Eq.(\ref{eq:Sent})  even for a single disorder realization, and even for relatively small systems. 

\begin{figure}[t]
\begin{center}
\includegraphics[width=\columnwidth]{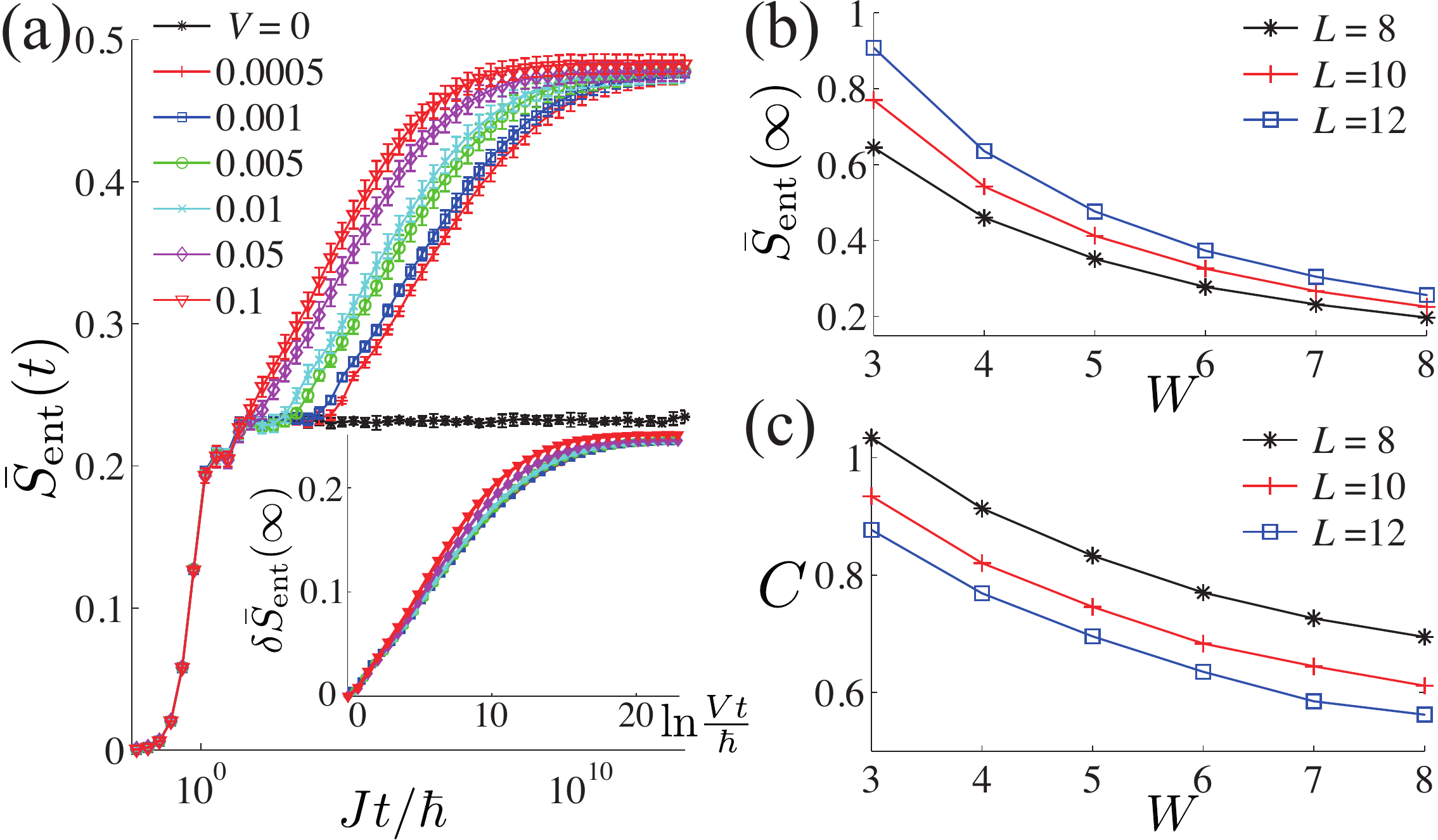}
\caption{ 
\label{Fig:updown} (Color online)
(a) Averaged entanglement entropy of initial product states, in which all fermions are localized at some sites, shows a characteristic logarithmic growth on long time scales (system size is $\Ss=12$, $\Ww  = 5$). Growth rate is found to be proportional to $\ln (Vt/\hbar)$ (inset). Saturated entanglement (b) and the ratio $C =\bar S_\text{\rm ent}(\infty)/\bar S_\text{diag}$ (c) decrease with $\Ww$ (for fixed $V=0.01$). 
}
\end{center}
\end{figure}
{\bf Numerical simulations.} In order to illustrate the above mechanism, and to explore the growth of entanglement for different initial states, we performed numerical simulations of the model~(\ref{eq:hamiltonian}) with a finite number of particles in a random potential uniformly distributed in the interval $[-W,W]$. Hopping is set to $J=1/2$ and we consider chains with an even number of sites and open boundary conditions at half filling.  The number of different disorder realizations ranged from 30\,000 ($L= 8$) to 800 ($L=14$); the number of initial states was $2^{L/2}$ for each disorder realization. In the figures below, error bars (if not shown) are approximately equal to the size of the symbols in each plot.

Using exact diagonalization for systems up to $14$ sites, we compute the time evolution of various initial states, which allows us to obtain $S_\text{\rm ent}(t)$ for half partition. We study its average $\bar S_\text{\rm ent}(t)$ over different initial states belonging to the same class, and over different realizations of disorder. 
Similar to Ref.~\cite{Moore12}, we first consider a class of localized product states (LPS) where each fermion is initially located at a given site.  Results for $\bar S_\text{\rm ent}(t)$ for a system of $\Ss=12$ sites and disorder $\Ww = 5$ are shown in Fig.~\ref{Fig:updown}(a).  After a rapid increase of entropy on time scales of the inverse hopping, due to diffusive transport on a scale smaller than the localization length, $\bar S_{\rm ent}(t)$ saturates for a noninteracting system. In the presence of even weak interactions, $\bar S_\text{ent}(t)$ continues to grow further. In full agreement with our analysis above, values of $\delta \bar S_{\rm ent} \equiv \bar S_\text{\rm ent}- \bar S_0$ collapse onto a single curve as a function of $\ln(V t/\hbar)$ [see the inset of Fig.~\ref{Fig:updown}(a)], where $\bar S_0$ is the saturation entropy of a noninteracting system.  

\begin{figure}[t]
\begin{center}
\includegraphics[width=\columnwidth]{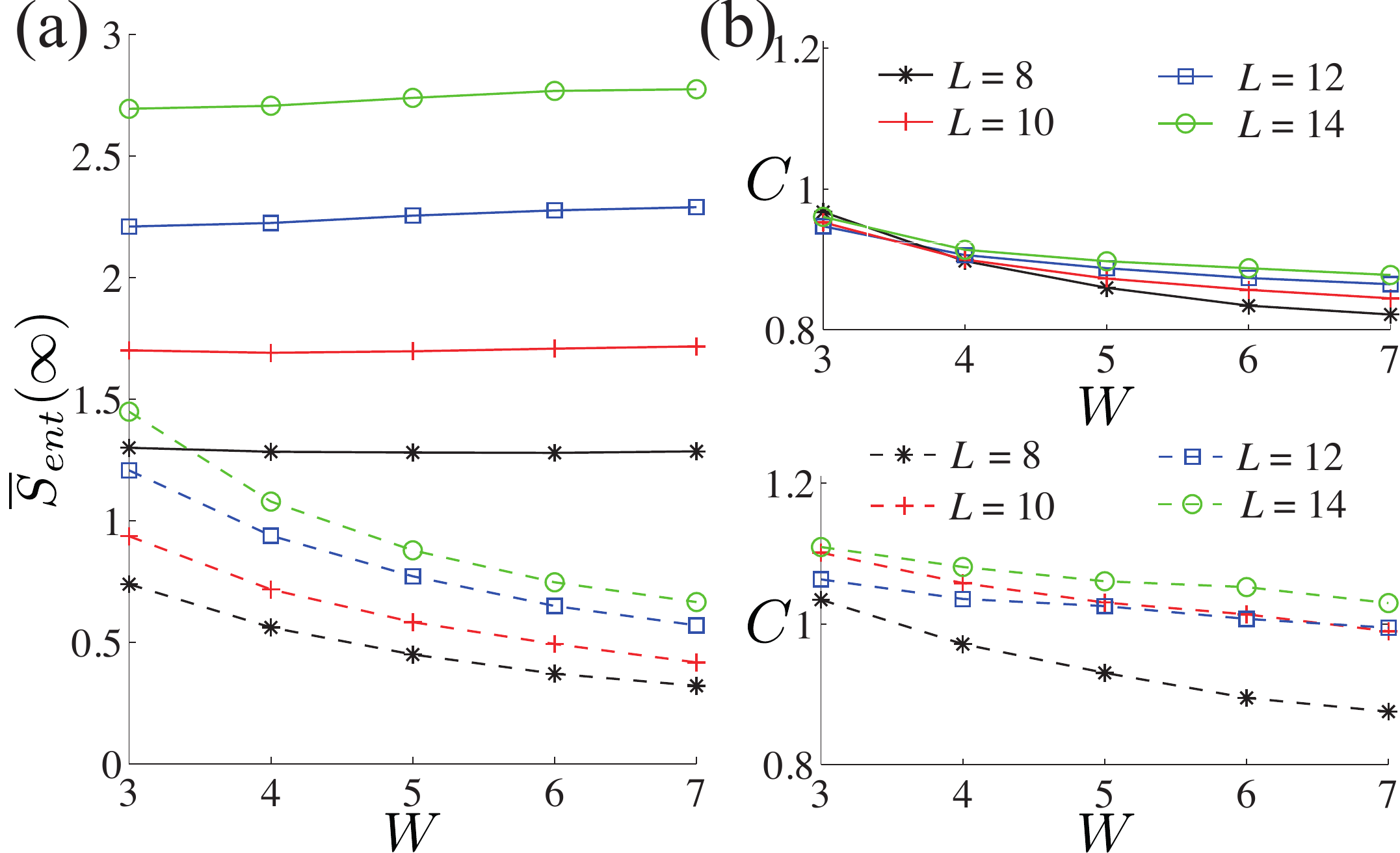}
\caption{ 
\label{Fig:other} (Color online) (a) Saturated entanglement entropy as a function of disorder $\Ww$ for a strongly entangled state (solid lines),  and a product of a strongly entangled state and an LPS state (dashed lines). (b) Ratio of saturated and diagonal entropy as a function of disorder $\Ww $ for the same two states. For the product of a strongly entangled state and an LPS state (lower panel), $C$ tends to 1 for larger system sizes as $W$ is increased (interaction is set to $V=0.01$).}
\end{center}
\end{figure}

The saturation value $\bar S_\text{ent}(\infty)$ does not vary appreciably with interaction strength when interactions are weak. This further supports the conclusion that weak interactions only weakly alter the eigenstates of the system. For fixed $V=0.01$,  $\bar S_{\rm ent}(\infty)$ and $S_{\rm diag}$ decrease with disorder~[see Fig.~2(b)] approximately as $1/W$ (scaling not shown). Such scaling stems from the fact that when $\xi$ is of the order of one lattice spacing, the leading contribution to the entanglement comes from rare resonant pairs of neighbouring sites (rather than typical off-resonant sites), which occur with probability $J/W$. Each pair contributes a number of the order $\sim \ln 2$ to the entanglement as well as diagonal entropy. 

We compare the saturated entanglement to $S_\text{diag}$, calculated  using the values of $P_i(\Ss/2)$ for the \emph{initial state} of the $\LS$ subsystem. The $P_i(\Ss/2)$ are obtained from the density matrix, using the eigenstates of the interacting Hamiltonian restricted to $\LS$. In this sense, while  $\bar S_\text{ent}(\infty)$ is determined from the  time evolution of the system,  $S_\text{diag}$ is solely the property of the initial state. 

To interpret the dependence of the ratio $C = \bar S_\text{ent}(\infty)/\bar S_\text{diag}$ on system size and disorder, Fig.~\ref{Fig:updown}(c), we must take into account two additional effects important for small systems: (i) the diffusion of particles across the entanglement cut; and (ii) the inefficiency of  decoherence when the number of terms in Eq.~(\ref{eq:initial_state}) is small or when $\LS$ and $\RS$ are of equal size.  These effects counteract each other, as diffusion leads to an additional contribution to $S_{\rm ent}$ not captured by Eq.~(\ref{eq:sinf}). On the other hand, inefficient decoherence leads to incomplete dephasing, and decreases $\bar S_\text{ent}(\infty)$ compared to $S_\text{diag}$. The positive contribution from (i) is  suppressed for larger systems or smaller localization lengths. The effect of (ii) depends on the initial state. For LPS  in the localized phase, the participation ratio is of order unity, and the effect (ii) is very pronounced. Thus,  $C$ is smaller than 1, and it decreases with increasing disorder or system size [see Fig.~\ref{Fig:updown}(c)].  

Next, we consider a different kind of initial states with larger participation ratios. The initial state of the $\RS$ and $\LS$ subsystem is chosen as a projection to the half-filled sector of the state $\prod_{i}\frac{1}{\sqrt{2}}(1 \pm c^\dagger_i) |0\rangle$, with $\pm$ signs chosen at random. Particles within each subsystem are therefore strongly entangled, but there is no entanglement between the subsystems at $t=0$. In this case, we find the same logarithmic entanglement growth, but $\bar S_{\rm ent}(\infty)$ [see the solid lines in Fig.~\ref{Fig:other}(a)] is larger compared to the previous case, and varies weakly with disorder. The ratio $C$ [see the upper panel of Fig.~\ref{Fig:other}(b)] now scales to 1 when system size is increased, contrary to the LPS. For this type of initial state, due to larger values of $\bar S_\text{ent}(\infty)$, the boundary diffusion contribution is less important; also, the superposition of a large number of eigenstates in each half of the system makes decoherence more efficient; thus, $C$ is closer to 1.  

Finally, we construct an example where $S_{\rm ent}$ reaches $S_{\rm diag}$. We take a product of the LPS  in $\LS$, and the strongly entangled state  in $\RS$. To further suppress the diffusion, we require the two sites adjacent to the entanglement cut to be always empty. $\bar S_\text{ent}(\infty)$ displays the behavior similar to the LPS case [see the dashed lines in Fig.~\ref{Fig:other}(a)], but is  larger due to the more effective dephasing. Remarkably, Fig.~\ref{Fig:other}(b) demonstrates that for larger system sizes, saturation and diagonal entropies become equal, in agreement with the above analysis. 

{\it Discussion.---}To summarize, we presented a mechanism of the logarithmic growth of entanglement in the MBL phase. We also established the laws governing the entanglement growth, and tested them in numerical simulations for different initial states. We note that in the delocalized phase the entanglement is expected to grow much faster (as a power-law function of time), suggesting that the scaling of $S_{\rm ent}$ can be used as a potential tool for studying the localization-delocalization critical point and its properties.

Although we focused on the limit of weak interactions, in which the eigenstates are similar to those of a noninteracting model, we expect our conclusions to hold also for stronger interactions which do modify the eigenstates. In this case, $\bar S_{\rm ent}(\infty)$ is expected to be determined by the participation ratios of the initial state in the basis of the interacting subsystem's eigenstates. Furthermore, our conclusions are expected to apply to localized interacting systems in any number of spatial dimensions. 

Our work indicates an exponentially broad distribution of dephasing time scales present in a MBL system. It gives support to the ``strong-localization" scenario of the many-body localization transition, and shows that the entanglement growth arises due to interaction-induced dephasing, rather than due to the effect of interactions on the eigenstates, as was hypothesized in Ref.~\cite{Moore12}. 

We note that recently Vosk and Altman~\cite{Altman13} considered an $XXZ$ model with random exchange interactions, but without a random field. For a special initial state, they developed a strong-disorder renormalization group procedure, and found that $S_{\rm ent}$ grows as a power of $\ln t$. The difference from our result stems from the fact that the state considered in Ref.~\cite{Altman13} was critical; however, the basic underlying mechanism---dephasing due to exponentially weak interactions between remote spins---is qualitatively similar. After this Letter was submitted, we became aware of a related independent work~\cite{Huse13} where the logarithmic growth of entanglement was established from phenomenological considerations.

{\it Acknowledgments.---}We would like to thank E. Altman and J. Moore for useful comments on the manuscript.  This research was supported in part by Perimeter Institute for Theoretical Physics. Research at Perimeter Institute is supported by the Government of Canada through Industry Canada and by the Province of Ontario through the Ministry of Economic Development \& Innovation. Z.P. was supported by DOE Grant No. DE-SC0002140. The simulations presented in this article were performed on computational resources supported by the High Performance Computing Center (PICSciE) at Princeton University.

\end{document}